\documentclass[conference]{IEEEtran}
\usepackage[utf8]{inputenc}
\IEEEoverridecommandlockouts

\IEEEoverridecommandlockouts

\usepackage{threeparttable, booktabs}
\usepackage{relsize}
\usepackage{makecell}
\usepackage{cite}
\usepackage{amsmath,amssymb,amsfonts}
\usepackage{algorithmic}
\usepackage{graphicx}
\usepackage{textcomp}
\usepackage{xcolor}
\usepackage{soul}  
\usepackage{booktabs}		

\usepackage[T1]{fontenc}

\usepackage{svg,subfigure}
\def\BibTeX{{\rm B\kern-.05em{\sc i\kern-.025em b}\kern-.08em
    T\kern-.1667em\lower.7ex\hbox{E}\kern-.125emX}}

\usepackage{xcolor} 
\newcount\Comments  
\usepackage{comment}
\newcommand{\kibitz}[2]{\ifnum\Comments=1{\color{#1}{#2}}\fi}
\newcommand{\ma}[1]{\kibitz{black}{#1}}

\usepackage{eurosym}
    
\begin{document}

\title{Comparative Techno-economic Assessment of Wind-Powered Green Hydrogen Pathways\\

}
\author{\IEEEauthorblockN{Merlinda Andoni$^{1}$, Benoit Couraud$^{1}$, Valentin Robu$^{2,3}$, Jamie Blanche$^{1}$, Sonam Norbu$^{1}$, \\
Si Chen$^{1}$, Satria Putra Kanugrahan$^{1}$, David Flynn$^{1}$}
	
	\IEEEauthorblockA{$^{1}$James Watt School of Engineering, University of Glasgow, Glasgow (UK)\\
        $^2$ CWI, Amsterdam (The Netherlands), $^3$ Eindhoven University of Technology (The Netherlands)\\
        }
}

\maketitle

\begin{abstract}
Amid global interest in resilient energy systems, green hydrogen is considered vital to the net-zero transition,
~yet its deployment remains limited by high production cost.
~The cost is determined by the its production pathway, system configuration, asset location, and interplay with electricity markets and regulatory frameworks. To compare different deployment strategies in the UK, we develop a comprehensive techno-economic framework based on the Levelised Cost of Hydrogen (LCOH) assessment. We apply this framework to 5 configurations of wind-electrolyser systems, identify the most cost-effective business cases, and conduct a sensitivity analysis of key economic parameters. Our results reveal that electricity cost is the dominant contributor to LCOH, followed by the electrolyser cost. Our work highlights the crucial role that location, market arrangements and control strategies among RES and hydrogen investors play in the economic feasibility of deploying green hydrogen systems. Policies that subsidise low-cost electricity access and optimise 
deployment can 
lower LCOH, enhancing the economic competitiveness of green hydrogen.
\end{abstract}


\section{Introduction}
The climate emergency and ambitious net-zero targets  position hydrogen as a key alternative to replace gas and fossil fuels,
~especially in 
difficult to decarbonise sectors via electrification, such as aviation, freight transportation, and heavy industry~\cite{haugen2022electrification}. Many countries have established roadmaps and policy to support hydrogen, e.g. the UK aims to deploy
~10 GW of low-carbon hydrogen capacity by 2030~\cite{ukTarget}.%
~Carbon-free green hydrogen produced by water electrolysis and renewable energy sources (RES) is preferred for decarbonisation, yet it accounts for less than 1\% of the global hydrogen market, mainly due to
~its high cost compared to fossil fuel-based alternatives, and the complex integration of large-scale production, distribution, and storage into existing energy infrastructure and systems.
~Government support schemes aim to lower costs, but they typically require proof of a low-emission production process, such as using RES or low-carbon grid electricity. For example in the UK, a project needs to demonstrate compliance with the Low Carbon Hydrogen Standard (LCHS), i.e. it needs to produce less than 50 gCO2e per kWh of electricity input or hold Renewable Energy Guarantees of Origin (REGOs).

RES deployment underpins the adoption of green hydrogen; however, globally and in the UK, RES face long connection queue backlogs and are responsible for growing costs related to curtailment.
~Hydrogen production may offer an 
alternative pathway for RES projects awaiting grid access, while reducing curtailment costs. 
~However, there is a lack of understanding on how to evaluate its potential due to the variety of deployment options and business cases, which include different physical configurations, commercial arrangements, location, operation strategies, and interplay with electricity markets and regulatory frameworks.
~To address such complexity, we develop a comprehensive techno-economic framework
~based on a 10 MW reference wind-electrolyser system in the UK. We assess the viability of different system configurations
~by estimating the Levelised Cost of Hydrogen (LCOH).
~Specifically, we focus on different
~grid connection types, contractual arrangements between developers, control strategies, and use cases, including `behind-the-meter', using the grid as back-up, and utilisation of curtailment. This work aims to inform decision-making on identifying the most promising use cases for green hydrogen in the UK context, with the methodology being relevant for analysis in other regions. Finally, we conduct a sensitivity analysis on key LCOH parameters to assess its evolution under future scenarios.

The remainder of this paper is structured as follows: Section~\ref{sec:system} introduces the wind-electrolyser system under study, Section~\ref{sec:TE_analysis} details the techno-economic framework,
~Section~\ref{sec:use_cases} elaborates on the green hydrogen use cases,
~Section~\ref{sec:results} shows the results and the LCOH sensitivity analysis, while Section~\ref{sec:fut_work} concludes this work and suggests future directions.

\section{Wind-electrolyser system}
\label{sec:system}
\begin{figure}[tbp]
\centering
\includegraphics[width=0.85\columnwidth]{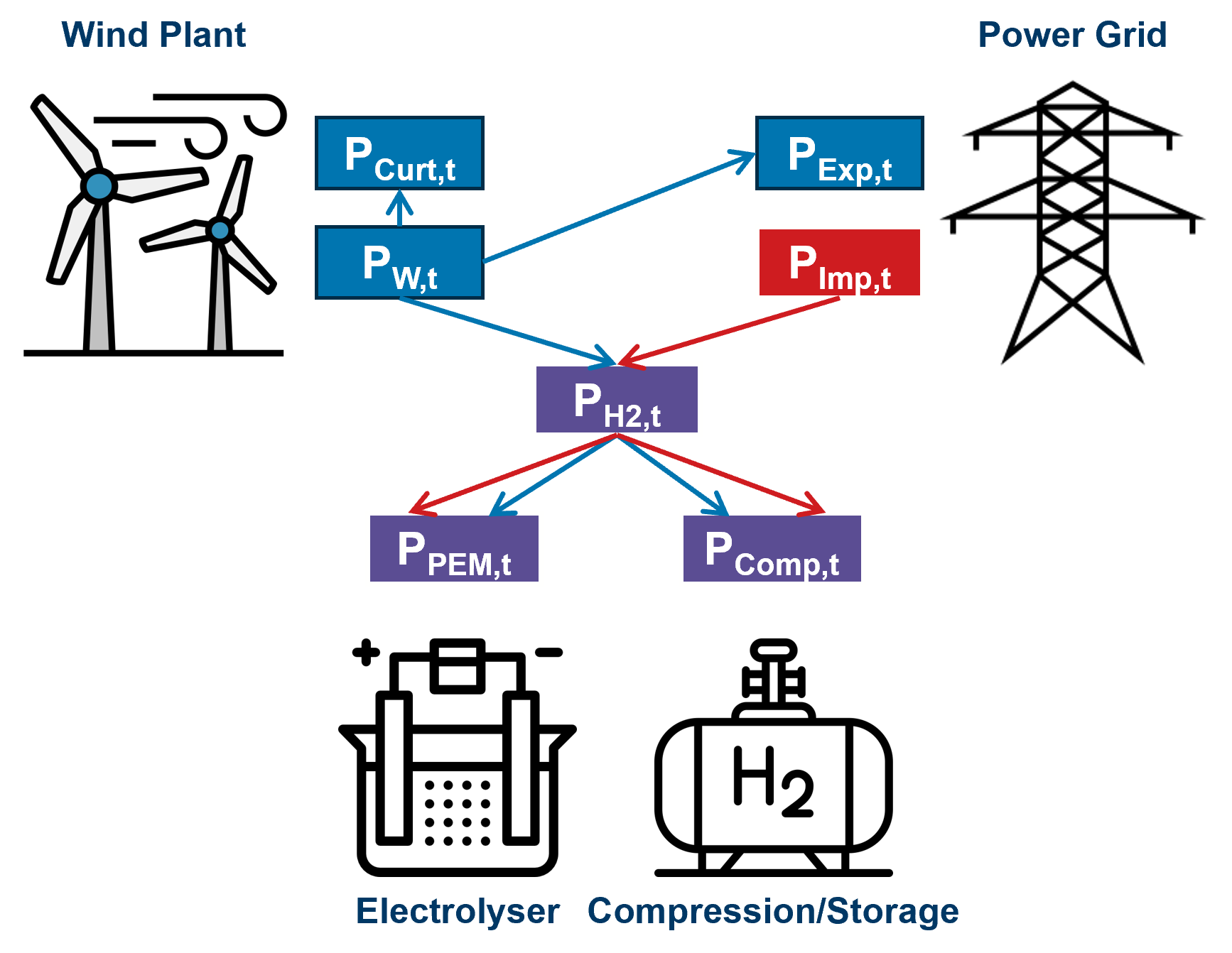}
\vspace{-5mm}
\caption{Wind-electrolyser system schematic illustrating potential power flows}
\vspace{-5mm}
\label{fig:power_flows}
\end{figure}

The analysis in this paper is based on a reference system comprising a 10 MW wind plant and a proton exchange membrane electrolyser (PEMEL) illustrated in Fig.~\ref{fig:power_flows}. PEMEL was selected because of its advantages compared to other technologies (alkaline and solid oxide), including high current density, fast response, and high operating pressure, resulting in fewer need for compression and storage, and its suitability for RES applications~\cite{vives2023techno}. 
Moreover, we assume a compressing unit, typically a mechanical compressor required to increase the volumetric energy density of produced hydrogen, and local storage, usually comprising above the ground metallic cylinders~\cite{vives2023techno}. 
Except for cases where hydrogen is consumed locally, it typically needs to be transported to end-use locations through pipelines or storage/distribution vehicles~\cite{christensen2020assessment}, or blended into the gas network. In this study, we exclude any transport and distribution costs and focus solely on the local compression and storage requirements at the production site.
Moreover,
~we consider wind-electrolyser systems of different forms, with wind turbines either co-located with the electrolyser or connected via the grid. The various possible power flow pathways between the wind plant, the PEM electrolyser (PEMEL), and the public grid are illustrated in Fig.~\ref{fig:power_flows} 
For every time step $t$, the wind output $P_{W,t}$ is allocated as:
\begin{equation}
    P_{W,t} = P_{H2,t} + P_{Exp,t} + P_{Curt,t}
\end{equation}
where $P_{H2,t}$ is the power used for hydrogen production,
~$P_{Exp,t}$ the power exported to the grid and $P_{Curt,t}$ the power curtailed, due to restricted grid access or network constraints. $P_{H2,t}$ is used to supply both the electrolyser and the compressor/storage system and may be supplied by either or a combination of the wind generator $P_{W,t}$ and grid imports
~$P_{Imp,t}$.
~Since the compressor's power demand depends on the hydrogen production rate, $P_{H2,t}$ is split into electrolysis $P_{PEM,t}$ and compression/storage $P_{Comp,t}$ by an iterative process: first, all available power is assigned to the electrolyser and we estimate the resulting hydrogen output, next we compute the energy required for compression, deduct it from $P_{PEM,t}$, and recalculate the PEMEL production. The process is repeated until the electrolysis/compression allocation
~converges. For every $t$, the hydrogen output is a function of the PEMEL input power and efficiency~\cite{buttler2018current}:
\begin{equation}
    M_{H_{2},t} = \frac{P_{PEM,t}\cdot\Delta t\cdot\eta(P_{PEM,t})}{LHV}
\end{equation}
where $\Delta t$ is the interval between two consecutive time steps ($0.5h$ in our case) and the energy requirement is based on the Lower Heating Value $LHV = 33.3$ kWh/kg. The next section, describes the techno-economic analysis framework.

\section{Techno-economic analysis framework}
\label{sec:TE_analysis}

\begin{table}[t!]
  \centering
  \resizebox{\columnwidth}{!}{%
  \begin{threeparttable}
    \caption{Summary of assumptions for techno-economic analysis}
    \label{tab:assumptions}
    \small
    \renewcommand{\arraystretch}{1.1}
    \begin{tabular}{llclc}
      \toprule[\heavyrulewidth]\toprule[\heavyrulewidth]
      \textbf{Parameter}               & \textbf{Notation}             & \textbf{Value}     & \textbf{Unit}                 & \textbf{Ref.} \\
      \midrule
      Wind capacity                    & $P_{W,r}$                     & 10                 & MW                            &  \\
      Lifetime wind                    & $n_{W}$                       & 25                 & years                         &  \\
      CAPEX wind                       & $CAPEX_{W}$                   & 1,230              & \pounds/kW                    & \cite{desnez_gen_costs} \\
      OPEX wind                        & $OPEX_{W}$                    & 25.4               & \pounds/kW/year               & \cite{desnez_gen_costs} \\
      Electrolyser capacity            & $P_{PEM,r}$                   & [5,10]\tnote{a}    & MW                            &  \\
      Lifetime electrolyser            & $n_{PEM}$                     & 30                 & years                         &  \\
      CAPEX electrolyser               & $CAPEX_{\mathsmaller{PEM}}$                 & 1,500\tnote{b}              & \pounds/kW                    &  \\
      OPEX electrolyser                & $OPEX_{\mathsmaller{PEM}}$                  & 5                  & \%\,CAPEX/year       & \cite{desnez_gen_costs} \\
      Stack lifetime                   & $n_{S}$                       & 60{,}000           & hours                         &  \\
      Stack replacement                & $C_{\text{replace},S}$        & 48                 & \%\,CAPEX           & \cite{vives2023techno} \\
      CAPEX compression                & $CAPEX_{\mathsmaller{Comp}}$         & 2.49               & \pounds/kg\tnote{c}    & \cite{penev2019economic} \\
      OPEX compression                 & $OPEX_{\mathsmaller{Comp}}$          & 6                  & \%\,CAPEX/year & \cite{penev2019economic} \\
      Energy compression               & $E_{\mathsmaller{Comp}}$             & 0.399              & kWh/kg                        & \cite{penev2019economic} \\
      Line capacity    & $P_{Line}$                        & [0,5,8,10]\tnote{d} & MW                           &  \\
      Wire capacity          & $Wire$                        & 0–10               & MW                            &  \\
      Grid access cost                 & $C_{Line}$                    & 100                & \pounds/kW                    &  \\
      Private wire cost                & $C_{Wire}$                    & 15                 & \pounds/kW                    &  \\
      Grid price (retail)       & $p_{\text{imp},r}$            & 0.184              & \pounds/kWh                   & \cite{desnez_prices} \\
      Wind–PEM price            & $p_{PPA}$                     & 0.057              & \pounds/kWh                   & \cite{vives2023techno} \\
      Grid export price                & $p_{exp}$                     & [0.07,0.044]\tnote{e}               & \pounds/kWh                   &  \\
      Discount rate                    & $d$                           & 3                  & \%                            &  \cite{vives2023techno}\\
      \bottomrule[\heavyrulewidth]
    \end{tabular}

    \begin{tablenotes}
      \footnotesize
      \item[a] Electrolyser capacity is 10MW in all use cases except Use Case V (5MW).
      \item[b] Conservative view of values reported in the literature (range of \$650-\$2,500/kW observed depending on scale, inclusion of balance of plant (BoP) and installation costs).
      \item[c] Of production capacity.
      \item[d] Line capacity: I \& III = 10MW; II = 0MW; IV = 8MW; V = 5MW  
      \item[e] $p_{exp}=0.07$ everywhere except for Use Case V-b cases where $p_{exp}=0.044$.
    \end{tablenotes}
  \end{threeparttable}
  }
\end{table}
This study estimates the LCOH to evaluate the economic viability of green hydrogen. The LCOH accounts for capital expenditures (CAPEX), fixed and variable operating costs (OPEX), electricity costs, and interconnection costs (grid access and/or any private wire connecting the PEMEL and wind plant). The LCOH is assessed on an annual half-hourly simulation basis.
~Lifetime asset cost components
~are annualised by multiplication with the capital recovery factor (CRF) that converts expenditure into an equivalent annual cost:
\vspace{-2mm}
\begin{equation}
    CRF(d,n) = \frac{d(1+d)^n}{(1+d)^n-1}
\end{equation}
where $d$ is the discount rate and $n$ the asset lifetime in years.
    \begin{figure*}[!h]
\vspace{-3mm}
\centerline{\includegraphics[width=0.9\textwidth]{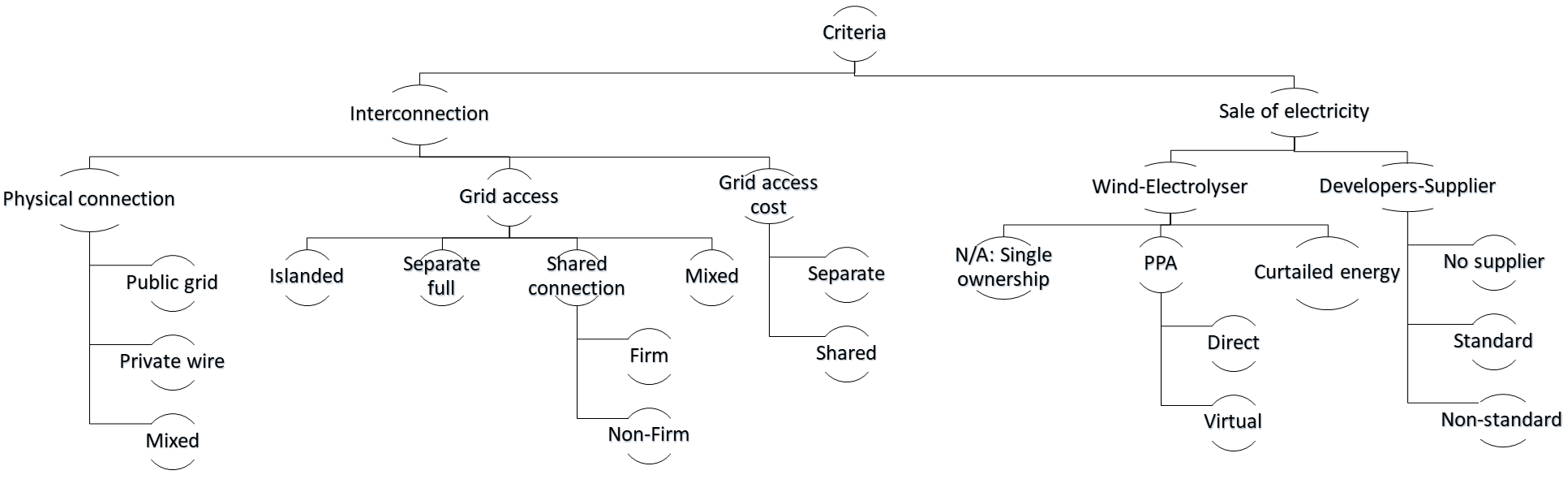}}
	\caption{Key features considered when conceptualising potential use cases}
    \vspace{-3mm}
	\label{fig:Criteria}
\end{figure*}
~The LCOH is estimated as the cost per mass unit of hydrogen produced ($\pounds/kg$) or as the cost per energy unit of hydrogen in (£/MWh H2 HHV: High Heating Value = 39.4 kWh/kg):
\vspace{-2mm}
\begin{equation}
    LCOH = \frac{C_{\mathsmaller{tot}}^{\mathsmaller{(\alpha)}}}{M_{\mathsmaller{{H_2}}}^{\mathsmaller{(\alpha)}}} \quad \mathrm{or} \quad\frac{C_{\mathsmaller{tot}}^{\mathsmaller{(\alpha)}}}{M_{\mathsmaller{{H_2}}}^{\mathsmaller{(\alpha)}} \times HHV /1,000}
\end{equation}
\vspace{-2mm}
where   $M_{{H_{2}}}^{{(\alpha})} = \displaystyle{\sum_t M_{{H_{2},t}}}$  is the annual hydrogen production and $\displaystyle{C_{{tot}}^{{(\alpha)}}}$ is the total annual cost equal to the sum of the electrolyser cost $C_{\mathsmaller{PEM}}^{\mathsmaller{(\alpha)}}$, the compressor cost $C_{\mathsmaller{Comp}}^{\mathsmaller{(\alpha)}}$, the cost of interconnection $C_{\mathsmaller{IC}}^{\mathsmaller{(\alpha)}}$ and the cost of electricity $C_{\mathsmaller{Elec}}^{\mathsmaller{(\alpha)}}$: 
\begin{equation}
    C_{\mathsmaller{PEM}}^{\mathsmaller{(\alpha)}} = P_{\mathsmaller{PEM,r}}\times(CAPEX_{\mathsmaller{PEM}}^{\mathsmaller{(\alpha)}} + OPEX_{\mathsmaller{PEM}}^{\mathsmaller{(\alpha)}})
\end{equation}
where:
$CAPEX_{\mathsmaller{PEM}}^{\mathsmaller{(\alpha)}} = (CAPEX_{\mathsmaller{PEM}}+LC_{\mathsmaller{replace,S}})\times CRF_{\mathsmaller{PEM}}$, $LC_{\mathsmaller{replace,S}}=\displaystyle{\frac{C_{\mathsmaller{replace,S}}}{(1+d)^{n_{S,y}}}}$ is the levelised stack replacement cost and $n_{S,y}$ the stack lifetime in years computed as the ratio of the stack lifetime in hours $n_S$ divided by the equivalent full load hours of the electrolyser equal to $\displaystyle{\sum_t \frac{{P_{\mathsmaller{PEM,t}}}\Delta t}{P_{\mathsmaller{PEM,r}}}}$. The compressor costs are:
\begin{equation}
    C_{\mathsmaller{Comp}}^{\mathsmaller{(\alpha)}} = CAPEX_{\mathsmaller{Comp}}^{(\mathsmaller{\alpha)}} + OPEX_{\mathsmaller{Comp}}^{\mathsmaller{(\alpha)}}
\end{equation}
where $CAPEX_{\mathsmaller{Comp}}^{\mathsmaller{(\alpha)}} = CAPEX_{\mathsmaller{Comp}}\times \mathrm{Cap}_{\mathsmaller{H_{2}}}^{(\mathsmaller{\alpha)}} \times CRF_{\mathsmaller{Comp}}$ and $\mathrm{Cap}_{\mathsmaller{H_{2}}}^{\mathsmaller{(\alpha)}}$ is compressor's hydrogen capacity, expressed in kg/year.
The interconnection costs are given by:
\begin{equation}
    C_{\mathsmaller{IC}}^{\mathsmaller{(\alpha)}} = P_{\mathsmaller{Line}}C_{\mathsmaller{Line}}^{\mathsmaller{(\alpha)}} + P_{\mathsmaller{Wire}}C_{\mathsmaller{Wire}}^{\mathsmaller{(\alpha)}}
\end{equation}
where $C_{\mathsmaller{Line}}^{\mathsmaller{(\alpha)}} = C_{\mathsmaller{Line}}\times CRF_{\mathsmaller{Line}}$ are the annualised costs related to the connection with the main grid and $C_{\mathsmaller{Wire}}^{\mathsmaller{(\alpha)}} = C_{\mathsmaller{Wire}}\times CRF_{\mathsmaller{wire}}$ the costs associated with the private wire.
~Finally, the electricity cost is:
\begin{equation}
    C_{\mathsmaller{Elec}}^{\mathsmaller{(\alpha)}} = {\sum_t P_{\mathsmaller{H2,W,t}}\Delta t} \times p_{\mathsmaller{PPA}} + {\sum_t P_{\mathsmaller{H2,Imp,t}}\Delta t} \times p_{\mathsmaller{imp,t}}
\end{equation}

The main assumptions are illustrated in Table~\ref{tab:assumptions}. The next section describes the use cases explored in this study.

\section{Green hydrogen use cases}
\label{sec:use_cases}

\begin{figure*}[ht!]
\vspace{-1mm}
	\centerline{\includegraphics[width=0.9\textwidth]{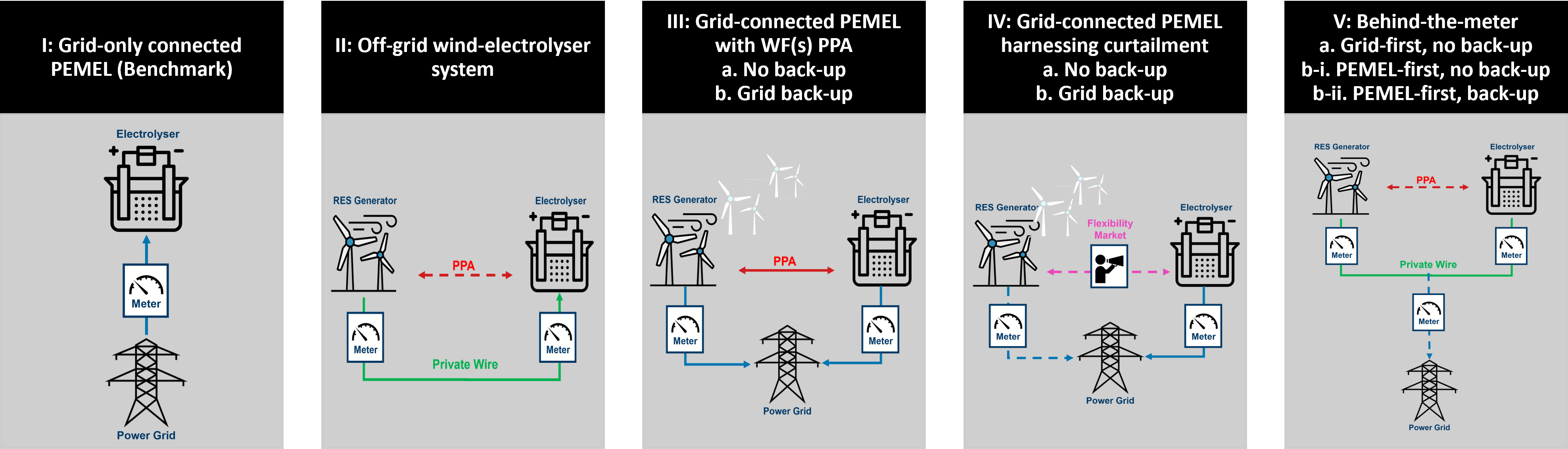}}
	\caption{Illustration of use cases assumed in this study}
    \vspace{-1mm}
	\label{fig:UseCases}
\end{figure*}

\begin{table*}[t!]
\centering
    \caption{Main Results of Comparative Analysis Across Use Cases}
    \label{tab:results}
    \resizebox{0.97\textwidth}{!}{
    \small
    \renewcommand{\arraystretch}{1.15}
    \begin{tabular}{lcccccc}
        \toprule[\heavyrulewidth]\toprule[\heavyrulewidth]
        \textbf{Use Case} & 
        \makecell{\textbf{LCOH}\\(\pounds/MWh)} & 
        \makecell{\textbf{LCOH}\\(\pounds/kg)} & 
        \makecell{\textbf{LCOH Free}\\(\pounds/MWh)} & 
        \makecell{\textbf{Annual}\\H\textsubscript{2} (t)} & 
        \makecell{\textbf{Stack}\\\textbf{Life (yrs)}} & 
        \makecell{\textbf{PEMEL Load}\\\textbf{Factor (\%)}} \\
        \midrule
        I. Grid-only PEMEL & \pounds263.62 & \pounds10.39 & – & 1,760.58 & 6.88 & 100.00\% \\
        II. Off-grid Wind–PEMEL & \pounds128.55 & \pounds5.06 & \pounds58.92 & 859.65 & 14.56 & 48.83\% \\
        III-a. Grid-connected PEMEL–Wind PPA (no back-up) & \pounds130.22 & \pounds5.13 & – & 859.65 & 14.56 & 48.83\% \\
        III-b. Grid-connected PEMEL–Wind PPA (grid back-up) & \pounds186.93 & \pounds7.37 & – & 1,760.58 & 6.88 & 100.00\% \\
        IV-a. Grid-connected PEMEL–Curtailment (no back-up) & \pounds192.25 & \pounds7.57 & \pounds121.68 & 407.43 & 30.32 & 23.14\% \\
        IV-b. Grid-connected PEMEL–Curtailment (grid back-up) & \pounds226.30 & \pounds8.92 & \pounds209.97 & 1,760.58 & 6.88 & 100.00\% \\
        V-a. Behind-the-meter, Grid-first (no back-up) & \pounds156.35 & \pounds6.16 & \pounds85.94 & 288.17 & 21.48 & 32.74\% \\
        V-b-i. Behind-the-meter, PEMEL-first (no back-up) & \pounds117.34 & \pounds4.62 & \pounds46.55 & 559.00 & 11.01 & 63.50\% \\
        V-b-ii. Behind-the-meter, PEMEL-first (grid back-up) & \pounds162.27 & \pounds6.39 & \pounds117.32 & 880.29 & 6.88 & 100.00\% \\
        \bottomrule[\heavyrulewidth]
    \end{tabular}
    }
\end{table*}

The selection of use cases for green hydrogen production analysis requires a structured methodology to navigate the breadth of available options and their inherent complexities. In this study, we employ a comprehensive mapping exercise that enumerates all viable configurations in which an electrolyser utilises wind energy in the UK context, including aspects related to interconnection, potential sourcing of electricity and commercial arrangements.
On interconnections, wind and electrolyser plants may be directly connected through a private wire, indirectly through the public grid, or a combination of both. Regarding how the public grid is accessed, projects may not be connected (islanded/off-grid), each asset may have a full and separate grid connection, they might share a firm or non-firm connection, or can adopt a mixture of the configurations above. Arrangements on interconnection can influence the LCOH as they can determine if and how much an electrolyser is exposed to grid connection costs. Electricity can be supplied under commercial agreements between wind-electrolyser developers (e.g., PPAs or curtailed-energy contracts) or provided free in the case of a sole investor or use of curtailed energy.
~Moreover, there are agreements between the developers and an energy supplier, as illustrated in Fig.~\ref{fig:Criteria}.
For example, investors may be subject to fixed or dynamic electricity prices. Electricity tariffs provided by energy suppliers typically include system charges, i.e. network charges and environmental levies
~which can account for up to 45\% of total LCOH in the UK~\cite{splitting}. Different system configurations can determine exemptions for such charges or compliance with government support schemes, like LCHS in the UK, which generally improve access to financing and grants.
~After careful consideration of the criteria in Fig.~\ref{fig:Criteria}, 5 main use cases were identified, as described below.

\subsection{Use case I: Grid-only electrolyser (Benchmark)}
In this benchmark case, a 10 MW electrolyser draws all its power from the public grid under a standard agreement with an energy supplier offering typical electricity tariffs, which could be fixed, dynamic, based on time-of-use (ToU) or green energy (when purchase of electricity is accompanied by REGOs) that could be used for compliance with low-carbon certification standards. For the analysis, we assume a fixed price based on the latest published average industrial retail price as in~\cite{desnez_prices}.
~The detailed assumptions are illustrated in Table~\ref{tab:assumptions}.

\vspace{-1.5mm}
\subsection{Use case II: Off-grid wind-electrolyser system}
A 10 MW wind farm and 
~electrolyser are co-developed, co-located, and connected via a private wire. The assets are not grid-connected, eliminating energy supplier agreements and grid/system charges. Wind output is dedicated to green hydrogen production, hence qualifies under LCHS. A direct PPA governs the sales of wind production: $p_{\mathsmaller{PPA}}=0$ (free energy) for the case of a sole investor, or $p_{\mathsmaller{PPA}}=0.057$ \pounds/kWh as in~\cite{vives2023techno}, when the wind and PEMEL are owned separately.

\vspace{-1.5mm}
\subsection{Use case III: Grid-connected electrolyser with wind PPA}
The wind farm and PEMEL (10 MW each) are connected at separate grid points, hence an energy supplier is required. Wind output is dedicated to hydrogen production, although the physical delivery is achieved through the public grid (virtual PPA at $p_{\mathsmaller{PPA}}=0.057$ \pounds/kWh). We consider two scenarios:
(a) \emph{Wind-only}: the electrolyser runs exclusively on PPA-supplied wind energy
(b) \emph{Wind \& grid back-up}: shortfall up to 10 MW is met by energy imported from the grid at a price equal to the average industrial price $p_{\mathsmaller{imp,r}}=0.184$ \pounds/kWh, as in \cite{desnez_prices}.

\vspace{-1.5mm}
\subsection{Use case IV: Grid-connected electrolyser using curtailment}
A 10 MW electrolyser aims to utilise wind curtailment (otherwise wasted energy). For easier comparison with other uses cases, we assume 5 wind farms of 10 MW that experience curtailment beyond 8 MW, i.e. in total, the PEMEL can harness 2x5=10 MW of wind power. The PEMEL needs to be located at a grid point where increase of demand can absorb curtailment without violating any grid constraints. Curtailment can be purchased at a flexibility market run by the system operator.
~Similar to Use Case III, zero cost and $p_{\mathsmaller{PPA}}=0.057$ \pounds/kWh are considered and the scenarios:
~(a) \emph{Wind-only}: the electrolyser operates solely on curtailed wind energy; (b) \emph{Wind \& grid back-up}: shortfall met by grid imported power.


\subsection{Use case V: Behind-the-meter (partially) grid-connected wind-electrolyser system}
In this case, we assume a partial, non-firm grid connection of 5 MW, a 10 MW wind plant and a 5 MW PEMEL, which are co-located and connected through a private wire, the cost of which is split between developers. The wind-PEMEL system operates `behind-the-meter'. In addition to the grid back-up options ((i) Wind-only and (ii) Wind \& grid back-up) and $p_{\mathsmaller{PPA}}=0$ or $p_{\mathsmaller{PPA}}=0.057$ \pounds/kWh  explored in previous use cases, we also consider two dispatch priorities:
~(a) Grid-First rule: Wind exports to the grid at $p_{exp}=0.07$ \pounds/kWh up to 5 MW, and provides the surplus to the electrolyser. Grid access cost are borne by the wind developer.
~(b) Electrolyser-first rule: Wind feeds the PEMEL first (up to 5 MW) and excess is exported to the grid at $p_{exp}=0.044$ \pounds/kWh. Grid access costs are shared between the two investors.

\section{Results}
\label{sec:results}

\begin{figure}[tbp]
\centering
\includegraphics[width=0.88\columnwidth]{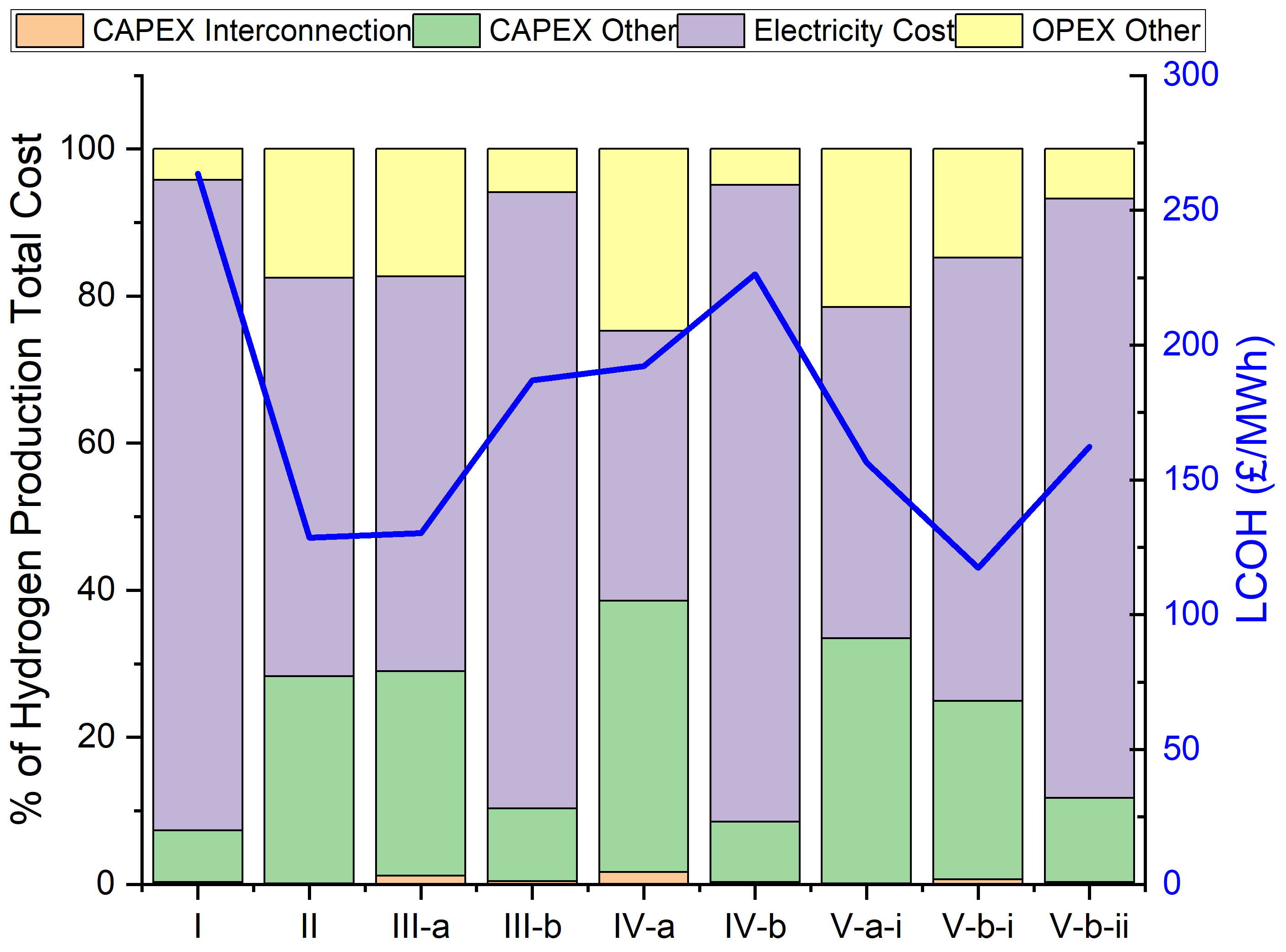}
\caption{Cost distribution as percentage of total cost of hydrogen production}
\vspace{-5mm}
\label{fig:costPercentage}
\end{figure}

The main results of the analysis are summarised in Table~\ref{tab:results}, \ma{where LCOH Free sets $p_{\text{PPA}}=0$ and serves as a benchmark for best-case scenarios.} The configuration achieving the lowest LCOH is use case V-b-i (behind-the-meter, electrolyser-first, without grid back up, followed by II (Off-grid wind-PEMEL system) and III-a (Grid-connected PEMEL with wind PPA). We find that no use case achieved a LCOH below \pounds100.00/MWh. The most expensive configuration is the grid-only electrolyser, and generally, in all cases with grid back-up the LCOH increases significantly, despite utilising the PEMEL at full load factor and achieving larger annual hydrogen production.
~The cost of electricity is the largest contributing factor across most use cases, as shown in Fig.~\ref{fig:costPercentage}, followed by PEMEL CAPEX costs. Use case IV, which utilises wind curtailment, is the second most expensive, due to the low utilisation of the PEMEL, which in turn increases the LCOH. In practice, LCOH depends on the level of curtailment that can be tolerated by wind developers (e.g. when running the simulations for 2 wind farms with a 5 MW partial connection each, the LCOH reduces to \pounds158.41/MWh or \pounds6.24/kg). In addition, if we assume pre-existing curtailment,
~free or low-cost use of curtailed energy is \ma{theoretically} plausible, indicating a potentially promising use case for green hydrogen production. \ma{In practice, curtailed energy access comes with a cost monetised via regulatory or contractual mechanisms, such as dedicated bilateral contracts for curtailed energy, participation in local flexibility markets, and system operator incentive schemes.} For the assumptions of Use Case IV, free electricity would lead to a LCOH of \pounds121.68/MWh. \ma{While reported costs present great variation due to specific cost and model assumptions, our findings are in line with recent techno-economic studies for green hydrogen production and wind-electrolyser systems outlined in Table~\ref{tab:LCOH_literature}, validating the realism of our modelling.}

\begin{figure}[!t]
\centering
\includegraphics[width=1\columnwidth]{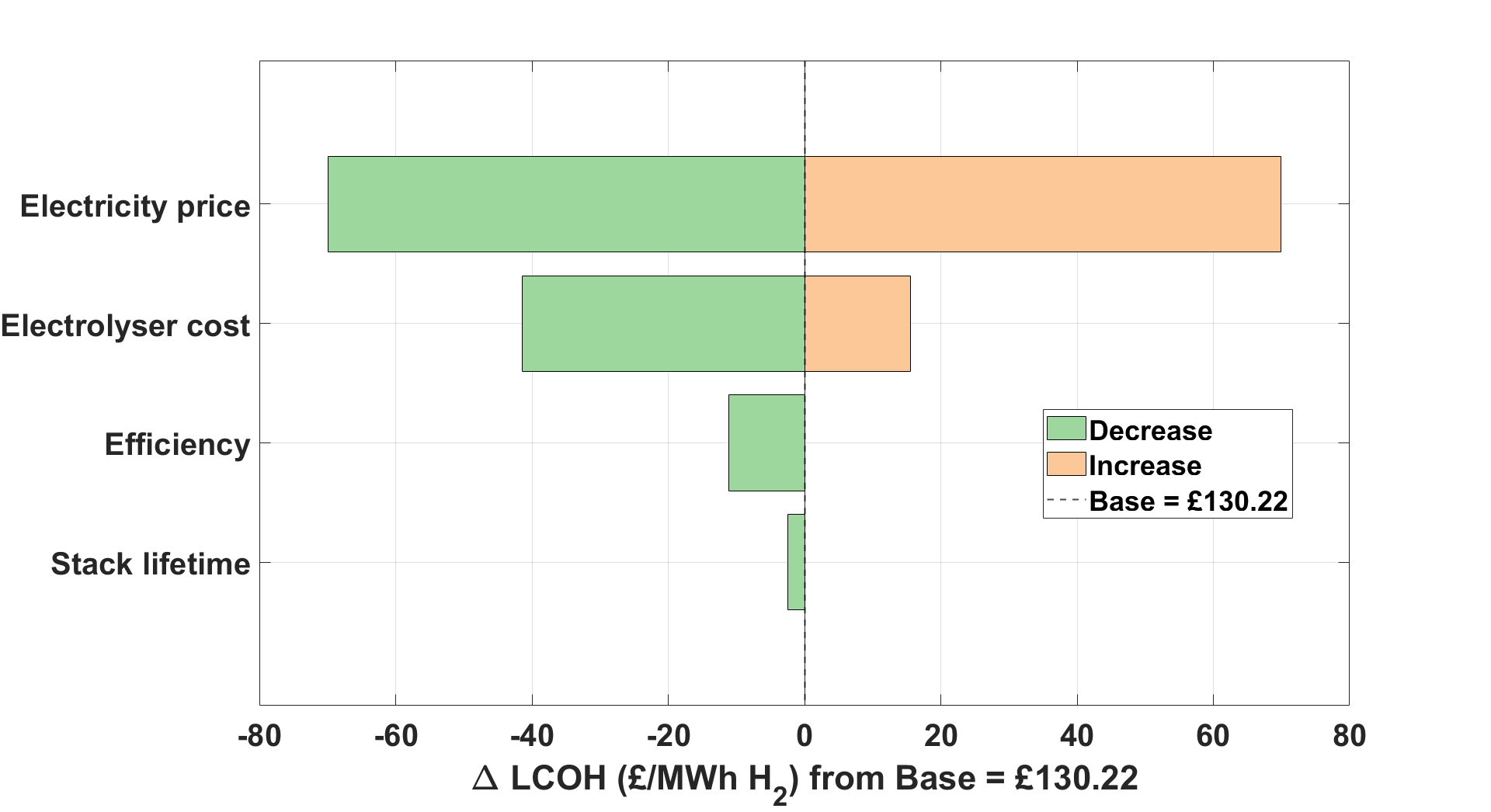}
\caption{Use case III-a: Sensitivity of LCOH: electricity price between investors ranges from [-100\%-+100\%], electrolyser cost [-80\%-+30\%], efficiency [+3\%, +5\%, +7\%, +10\%] and stack lifetime [60,000-120,000 hours]  }
\vspace{-5mm}
\label{fig:sensitivity}
\end{figure}

In addition, we perform a sensitivity analysis on the main parameters affecting the LCOH i.e. the electricity price (between developers and the grid electricity price), the PEMEL costs, future improvements on the electrolyser efficiency and lifetime, and discount rate to account for different economic parameters. Due to a lack of space and because the conclusions are similar across use cases, we indicatively show results for selected parameters in Use Case III-a in Fig.~\ref{fig:sensitivity}, where the central case refers to the assumptions in Table~\ref{tab:assumptions}.
\begin{table}[!t]
  \caption{\ma{Green hydrogen LCOH range reported in the literature}}
  \label{tab:LCOH_literature}
  \centering
  \begin{tabular}{|l|c|c|}
    \hline
    \textbf{Study} & \textbf{LCOH} & \textbf{Source} \\
    \hline
    Onshore/offshore wind \& solar (UK) & \pounds3.76-\pounds4.87/kg & \cite{de2025power} \\
    Offshore wind (UK) & \pounds9.08/kg (10.49\euro/kg) & \cite{hill2024cost} \\
    RES hydrogen (Hydrogen Council) & \$4.5-\$6.5/kg & \cite{hydrogen2023hydrogen} \\
    RES hydrogen (DOE US)& \$5-\$7/kg & \cite{hubert2024clean} \\
    Green hydrogen (IEA global) & \$3.4-\$12/kg & \cite{IEA2023global} \\
    \hline
  \end{tabular}

\end{table}

\section{Conclusions \& Future work}
\label{sec:fut_work}
This paper presented a techno-economic framework for the evaluation of 5 wind-electrolyser use cases. Our work highlights the crucial role that location, PPA arrangements and control strategies play in the economic feasibility of deploying green hydrogen systems. We show that behind-the-meter systems with an electrolyser-first rule and no grid back-up achieved the lowest LCOH. Co-located configurations also outperform set-ups using the public grid as back-up, while the grid-only electrolyser remains the most costly option. The viability of curtailment-based cases depends on the level of curtailment available and price. If curtailed energy can be purchased at low costs and even for free, use cases that utilise curtailed energy can be very promising. Across all use cases, the electricity price and the electrolyser cost emerge as the primary LCOH drivers.
~Overall, the attractiveness of green hydrogen would improve with subsidies that aim to reduce the electricity cost, network charges and environmental levies, while policy-makers should facilitate wind-electrolyser partnerships (co-located or virtually connected).

Future work will extend the framework by integrating co-optimising the system ratings,
~integration of real-time tariffs, and use of sophisticated operational strategies that aim to align hydrogen production with grid carbon intensity signals and certification schemes.
~\ma{Note that while our model results are based on onshore wind costs, the analysis also applies to offshore projects, which typically have higher capacity factors but also higher CAPEX and OPEX. Future work will explore co-located offshore wind-electrolyser systems, which may yield lower LCOH.} Directions of future work will equip investors and policymakers with insights on the deployment of cost-effective green hydrogen solutions, closing the gap to scalable adoption of green hydrogen.

\section*{Acknowledgment}
Supported by the EPSRC HI-ACT (“Hydrogen Integration for Accelerated Energy Transitions Hub”) [EP/X038823/2], DISPATCH ("DecarbonISation PAThways for Cooling and Heating") [EP/V042955/1], and DARe ("Decarbonised Adaptable and Resilient Transport Infrastructures") [EP/Y024257/1].

\bibliographystyle{ieeetr}
\bibliography{main}

\end{document}